\documentclass[12pt,preprint]{aastex}

\shorttitle{Resistive MHD accretion and jets}
\shortauthors{Kuwabara et al.}

\begin{document}


\title{THE ACCELERATION MECHANISM OF RESISTIVE MHD JETS LAUNCHED FROM
       ACCRETION DISKS}


\author{Takuhito Kuwabara}
\affil{Applied Research and Standards Department,
       National Institute of Information and Communications Technology,
       Koganei, Tokyo 184-8795, Japan}
\email{kuwabrtk@cc.nao.ac.jp}

\author{Kazunari Shibata}
\affil{Kwasan and Hida Observatories, Kyoto University, Yamashina,
       Kyoto 607-8471, Japan}
\email{shibata@kwasan.kyoto-u.ac.jp}

\author{Takahiro Kudoh}
\affil{Department of Physics and Astronomy, University of Western Ontario,
       London, Ontario N6A 3K7, Canada}
\email{kudoh@astro.uwo.ca}

\and

\author{Ryoji Matsumoto}
\affil{Department of Physics, Faculty of Science, Chiba University,
       Inage-ku, Chiba 263-8522, Japan}
\email{matumoto@astro.s.chiba-u.ac.jp}



\begin{abstract}

We analyzed the results of non-linear resistive magnetohydrodynamical
(MHD) simulations of jet formation to study the acceleration mechanism of
axisymmetric, resistive MHD jets. The initial state is a constant angular
momentum, polytropic torus threaded by weak uniform vertical magnetic fields.
The time evolution of the torus is simulated by applying the CIP-MOCCT scheme
extended for resistive MHD equations.
We carried out simulations up to 50 rotation period at the innermost radius
of the disk created by accretion from the torus.
The acceleration forces and the characteristics of resistive jets
were studied by computing forces acting on Lagrangian test particles.
Since the angle between the rotation axis of the disk and magnetic field
lines is smaller in resistive models than in ideal MHD models,
magnetocentrifugal acceleration is smaller.
The effective potential along a magnetic field line has maximum
around $z \sim 0.5r_0$ in resistive models, where $r_0$ is the radius
where the density of the initial torus is maximum.
Jets are launched after the disk material is
lifted to this height by pressure gradient force. Even in this case,
the main acceleration force around the slow magnetosonic point is the
magnetocentrifugal force. The power of the resistive MHD jet is comparable to
the mechanical energy liberated in the disk by mass accretion.
Joule heating is not essential for the formation of jets.

\end{abstract}




\section{INTRODUCTION}

Magnetically driven mass outflows from accretion disks have been studied
extensively. \citet{bla82} showed that a magneto-centrifugally
driven cold outflow emanates from an accretion disk when the angle $\theta$
between poloidal magnetic field lines threading the disk and the rotation
axis of the disk is larger than $30^{\circ}$. 
Nonlinear magnetohydrodynamical (MHD)
simulations of jet formation including accretion disks were first carried
out by \citet{uchida85} and \citet{shibata86}. Since the
magnetically driven jet extracts angular momentum from the disk, the jet
formation process enhances the accretion of the disk material. The back
reaction of the jet formation on disk accretion and its relation to the
magnetorotational instability \citep[MRI ;][]{bh91} were discussed
by \citet{stone94}, \citet{matsumoto96}, and \citet{kudoh02}.

A key question which is often raised to the time dependent simulations of
jet formation is whether the system approaches a steady state.
Axisymmetric ideal MHD simulations of jet formation including an
accretion disk show episodic outflows instead of approaching a quasi-steady
state \citep[e.g.,][]{kuwa00}. One successful approach to get steady jet
by non-steady simulations is to treat the disk as time-independent boundary
condition \citep*[e.g.,][for resistive MHD simulations]
{usty95,romanova97,ouyed97,fendt02}.
Another
approach is to include an accretion disk inside the computational box and
assume magnetic diffusivity
\citep{kuwa00,casse02,casse04}.

\citet{kuwa00} carried out 2.5-dimensional axisymmetric resistive
MHD simulations starting from a rotating torus initially threaded by weak
uniform vertical magnetic fields. They showed that the jet property
drastically changes depending on the magnetic diffusivity, which they
assumed to be uniform. When the magnetic diffusivity is small, mass
accretion and jet formation takes place intermittently. On the other hand,
in mildly diffusive disks, they showed by simulations for time scale
about 50 rotation period at the innermost radius of the disk that
both jets and accretion disks approach a quasi-steady state.
\citet{casse02,casse04} extended this study to the case of
geometrically thin disk initially threaded by equipartition
($\beta=P_{gas}/P_{mag} \sim 1$) poloidal magnetic fields. By carrying
out simulations assuming magnetic diffusivity localized inside the disk,
they achieved near stationary state. 

In order to study the
acceleration mechanisms and energy transport of resistive MHD jets more
quantitatively, we re-computed the simulation models adopted by \citet{kuwa00}
by applying the CIP-MOCCT scheme \citep{kudoh98} which is more robust
and accurate than the scheme we adopted in \citet{kuwa00}.

In section 2, we describe the assumptions and numerical methods. Numerical
results are presented in section 3. Discussions and conclusions are given
in section 4.

\section{MODELS}
\subsection{Assumptions and Basic Equations}
We solve two-dimensional nonlinear, time-dependent,
compressible resistive MHD equations in a cylindrical coordinate
system $(r,\, z)$ under the assumption of axisymmetry
to investigate the jet ejection from the accretion disk.
The $z$-direction is parallel to the rotational axis of the accretion disk
(see Fig. \ref{fig2}).
The basic equations are
\begin{equation}
\frac{\partial \rho}{\partial t}
+\mbox{\boldmath $V$}\cdot\mbox{\boldmath $\nabla$}\rho
=-\rho\mbox{\boldmath $\nabla$}\cdot\mbox{\boldmath $V$},
\label{eq:01}
\end{equation}
\begin{eqnarray}
\frac{\partial\mbox{\boldmath $V$}}{\partial t}
+\mbox{\boldmath $V$}\cdot\mbox{\boldmath $\nabla$}\mbox{\boldmath $V$}
\!\!\!\!&=&\!\!\!\!-\frac{1}{\rho}\mbox{\boldmath $\nabla$}P \nonumber \\
\!\!\!\!&+&\!\!\!\!\!\frac{1}{4\pi\rho}
\left(\mbox{\boldmath$\nabla$}\times\mbox{\boldmath$B$}\right)
\times\mbox{\boldmath$B$}
-\mbox{\boldmath $\nabla$}\psi,
\label{eq:02}
\end{eqnarray}
\begin{equation}
\frac{\partial\mbox{\boldmath$B$}}{\partial t}
   =\mbox{\boldmath$\nabla$}\times
\left(\mbox{\boldmath$V$}\times\mbox{\boldmath$B$}
      -\eta\mbox{\boldmath$\nabla$}\times\mbox{\boldmath$B$}\right),
\label{eq:03}
\end{equation}
\begin{eqnarray}
\frac{\partial P}{\partial t}
+\mbox{\boldmath$V$}\cdot\mbox{\boldmath$\nabla$}P
=\!\!\!\!&-&\!\!\!\!\!\gamma P\mbox{\boldmath$\nabla$}
 \cdot\mbox{\boldmath$V$} \nonumber \\
\!\!\!\!&+&\!\!\!\!\!\left(\gamma-1\right)\frac{\eta^{'}}{4\pi}
\left(\mbox{\boldmath$\nabla$}\times\mbox{\boldmath$B$}\right)^2,
\label{eq:04}
\end{eqnarray}
where $\psi$ is the gravitational potential
\begin{equation}
   \psi =-\frac{GM}{(r^2+z^2)^{1/2}}, 
\label{eq:05}
\end{equation}
$G$ is the gravitational constant, $M$ is the mass of
the central object, and
$\eta$ is the resistivity which is assumed to be uniform.
The other symbols have their usual meanings.
In equation (\ref{eq:04}), we set $\eta^{'}=\eta$ or $\eta^{'}=0$.
The latter corresponds to the case when we neglect Joule heating
(or equivalently, assume cooling which balances with the Joule heating).
The units of length, velocity, time, and density are
$r_0$, $V_{\rm K0}$, $r_0/V_{\rm K0}$, and $\rho_0$, where
$r_0$ is the radius where the density of the torus is maximum,
$V_{\rm K0}$, and $\rho_0$ are Keplerian rotation speed,
and the density at $(r,\, z)=(r_0,\, 0)$, respectively.
In this normalization,
we have two non-dimensional parameters:
\begin{eqnarray}
E_{\rm th}\!\!\!&=&\!\!\!\frac{V^2_{\rm s0}}{\gamma V^2_{\rm K0}},
   \label{eq:06} \\
E_{\rm mg}\!\!\!&=&\!\!\!\frac{V^2_{\rm A0}}{V^2_{\rm K0}},
   \label{eq:07}
\end{eqnarray}
where $V_{\rm s0}=(\gamma P_0/\rho_0)^{1/2}$,
and $V_{\rm A0}=[B_0^2/(4\pi\rho_0)]^{1/2}$
are the sound speed, and
Alfv\'en speed at $(r_0,\, 0)$, respectively.
Here, $E_{\rm th}$ is the ratio of thermal energy to gravitational energy
and $E_{\rm mg}$ is the ratio of magnetic energy to gravitational energy.
The normalized resistivity $\bar{\eta}$ is defined as 
$\bar{\eta}=\eta / (r_{\rm 0}V_{\rm K0})$ and the magnetic Reynolds
number at $(r_0,\, 0)$ is defined as 
$R_{\rm m0}\equiv r_0V_{\rm A0}/\eta$.

\subsection{Initial Condition}
We assume an equilibrium disk rotating around a central object
surrounded by a hot corona \citep[e.g.,][]{matsumoto96}.
The assumption of the existence of hot corona is a natural consequence of
energy transfer from magnetically active disks. Such corona exists
above galactic gas disks and solar photosphere 
\citep[see e.g.,][for accretion disk corona]{galeev81}.
Equilibrium solutions of a torus can be obtained
under the following simplifying assumptions.
Here, we adopted Newtonian analogue of Abramowicz's relativistic tori
\citep[][]{abramo78}.
The distributions of angular momentum is
\begin{equation}
   L=L_0\,r^a.       \label{eq:08}
\end{equation}
We assume polytropic equation of state
\begin{equation}
   P=K\,\rho^{1+1/n}.   \label{eq:09}
\end{equation}
The density distribution of the torus is determined by
\begin{eqnarray}
   -\frac{GM}{(r^2+z^2)^{1/2}}
   \!\!\!\!\!&+&\!\!\!\!\!\frac{1}{2(1-a)}L_0^2\,r^{2a-2} \nonumber \\
   \!\!\!\!\!&+&\!\!\!\!\!(n+1)\frac{P}{\rho}={\rm constant}. 
   \label{eq:10}
\end{eqnarray}
The mass distribution outside the torus is assumed to be that of
the isothermal non-rotating high temperature halo surrounding
the central object,
\begin{equation}
   \rho=\rho_{\rm h}\,\exp\!\left\{\alpha\left[\frac{r_0}{(r^2+z^2)^{1/2}}
   -1\right]\right\},   \label{eq:11}
\end{equation}
where $\alpha=\gamma V_{\rm K0}^2/V_{\rm sc}^2$. Here $V_{\rm sc}$
and $\rho_{\rm h}$
are the sound velocity and density in the halo
at $(0,\, r_0)$, respectively.
We assume that $a=0$ ($L=$ constant), $n=3$, $\gamma=5/3$, 
$\alpha =1.0$, $\rho_{\rm h}/\rho_0=10^{-3}$, 
$E_{\rm th}=5.0\times 10^{-2}$, and $E_{\rm mg}=5.0\times 10^{-4}$.
The initial magnetic field is assumed to be uniform and parallel to the
$z$-axis.

The magnetic Reynolds number is defined by $R_m = \lambda V_A/\eta
= R_{m0} (\lambda/r_0)(V_A/V_{A0})$, where
$\lambda = 2\pi V_A/\Omega$ is the characteristic scale of the
magnetorotational instability,
$V_A$ is the Alfv\'en velocity, and $\Omega$
is the angular velocity of the disk.
It increases from inside the torus to
the halo and it becomes $R_m \gg 1$ in the halo, thus magnetic diffusion
is not important there \citep{kuwa00}.

\subsection{Numerical Methods and Boundary Conditions}
We solved the equations (\ref{eq:01})--(\ref{eq:04}) 
by using the CIP-MOCCT method.
The algorithm of the original CIP-MOCCT method and results of test
simulations are described in \citet{kudoh98,kudoh99}.
In this method, the CIP scheme \citep{yabe91} is used for
hydrodynamical part and the MOCCT scheme \citep{stone92} is
used to solve the induction equation (\ref{eq:03}) and to evaluate the Lorentz
force terms. The basic equations (\ref{eq:01}), (\ref{eq:02}), and
(\ref{eq:04}) are expressed
in non-conservation form as
\begin{equation}
\frac{\partial f}{\partial t}+(\mbox{\boldmath $V$}\cdot 
\mbox{\boldmath $\nabla$})f = S
\end{equation}
where $S$ is the source term. We solve the equation in two steps;
advective step ${\partial f}/{\partial t}
+(\mbox{\boldmath $V$}\cdot\mbox{\boldmath $\nabla$})f=0$
and the source step ${\partial f}/{\partial t}=S$.
We revised the source step to ensure higher accuracy in space and time.
In the MOCCT step, we modified the original scheme such that we include
the resistive term in the induction equation and that characteristic
equation of Alfv\'en waves are solved by using the CIP scheme.

The size of simulation box is
$(r_{\rm max}\times z_{\rm max})=(5.1r_0\times 13.4r_0)$,
the number of grid points is $(N_r,\, N_z)=(200,\, 256)$,
and the grid size is $\Delta r=0.01r_0$ when $0\le r \le 1$
and $\Delta z=0.01r_0$ when $0\le z \le 1$.
Otherwise the grid size increases with $r$ and $z$.
At $r=0$,
we assume $\rho,\ P,\ V_z$, and $B_z$ are symmetric, while
$V_r, \ V_{\phi},\ B_r$, and $B_{\phi}$ are anti-symmetric.
We computed only the upper half plane ($z > 0$) by assuming
$\rho,\ P,\ V_r,\ V_{\phi},\ {\rm and}\ B_z $ are symmetric, while
$V_z,\ B_r,\ {\rm and}\ B_{\phi}$ are anti-symmetric 
with respect to $z=0$.
The outer boundaries at
$r=r_{\rm max}$ and $z=z_{\rm max}$ are free boundaries where waves can
be transmitted.
We softened the gravitational potential
inside $R=(r^2+z^2)^{1/2}=R_{\rm in}=0.2\,r_0$ 
to avoid the singularity at $R=0$.
As we show later, the disk material smoothly penetrates into the region
$R < R_{\rm in}$. $R_{\rm in}$ can be considered as the innermost radius of the
disk. The unit time $t_0=r_0/V_{K0} \sim 11R_{\rm in}/V_{\rm K,in}$ where
$V_{\rm K,in}$ is the Keplerian rotation speed at $r=R_{\rm in}$, corresponds
to 1.8 rotation period at $r=R_{\rm in}$.

We analyzed the results of three simulations (see Table \ref{tbl:1}).
Model $R$ is a mildly diffusive model 
($\bar{\eta}=\bar{\eta}^{'}=1.25\times 10^{-2}$).
Model $RC$ is the resistive model without Joule heating
($\bar{\eta}=1.25\times10^{-2},\,\bar{\eta}^{'}=0$).
Model $I$ is the non-diffusive model ($\bar{\eta}=\bar{\eta}^{'}=0$).
Model $R$ and Model $I$ are the same as those reported by \citet{kuwa00}.

To study the acceleration mechanism of the jet,
we put Lagrangian test particles near the disk surface and computed the time
evolutions of the location of these particles and evaluated the forces
acting on each particle as schematically shown 
in Figure \ref{fig2} ({\it left}).
The initial positions of particles
are selected such that they form main part of the jet.
In model $R$,
we put Lagrangian particles on a magnetic field line at $r=1.5\,r_0$,
between $z=0.64\,r_0$ and $0.84\,r_0$. In model $I$,
we put them on a magnetic field line at $r=0.8\,r_0$, 
between $z=0.31\,r_0$ and $0.46\,r_0$. 
The particle positions
are updated by using fluid velocity
$\mbox{\boldmath $V$}=(V_r,\ V_z)$ as follows:
$r_{\rm p}^{n+1}=r_{\rm p}^n+V_r\,\Delta t$,
$z_{\rm p}^{n+1}=z_{\rm p}^n+V_z\,\Delta t$.
Here, $r_{\rm p}$, and $z_{\rm p}$ are the position of particles and $n+1$,
and $n$
show the time step whose interval is $\Delta t$.
In Figure \ref{fig2} ({\it right}), we schematically show the force
${f}_{\rm p}^{'}$ 
defined as the projection
of poloidal force per mass $\mbox{\boldmath $f$}_{\rm p}$ 
to the direction of poloidal velocity
vector $\mbox{\boldmath $V$}_{\!\!\rm p}$ of the particle;
\begin{equation}
{f}_{\rm p}^{'}=|\mbox{\boldmath ${f}$}_{\!\rm p}|
\frac{\mbox{\boldmath ${f}$}_{\!\rm p}\cdot\mbox{\boldmath $V$}_{\!\!\rm p}}
     {|\mbox{\boldmath ${f}$}_{\!\rm p}||\mbox{\boldmath $V$}_{\!\!\rm p}|}.
\label{eq:12}
\end{equation}
When the sign of force ${f}_{\rm p}^{'}$ 
is plus, the force
accelerates the particle. On the other hand, when the sign of the force
${f}_{\rm p}^{'}$
is minus, it decelerates the particle, respectively.
By plotting the time variation of ${f}_{\rm p}^{'}$, 
we can check the acceleration force along the streamline
acting on particles.

\section{NUMERICAL RESULTS}
\subsection{Time Evolution of Jets}

Figure \ref{fig2.1} ({\it top panels})
show the time evolution of temperature distribution
(color scale), magnetic field lines (white curves) and velocity vectors 
(arrows) in model $R$. 
Numerical results are in good agreement with those reported by
Kuwabara et al. (2000). As Kudoh et al. (1998) have already shown,
numerical results obtained by the CIP-MOCCT scheme agree well with 
those obtained by the modified Lax-Wendroff scheme with
artificial viscosity, except that numerical oscillations in low-$\beta$
regions are not prominent when the CIP-MOCCT method is used 
and that contact surfaces are sharply traced with the CIP-MOCCT method.
Figure \ref{fig2.1} ({\it bottom panels}) show the time evolution of 
model $RC$.
In this model, resistivity is included in the induction equation 
but Joule heating term is not included in the energy equation.
Numerical results indicate that Joule heating is not essential for 
the jet formation. The collimation of the jet is better in the model
without Joule heating (model $RC$) because 
temperature of the jet is
lower than that in the model with Joule heating (model $R$).

Figure \ref{fig2.01} ({\it left}) shows the density distribution at $t=25$.
High density ridge is formed near the outermost radius of the initial torus.
This ridge is the contact surface between the disk material ejected from
the disk and 
the ambient halo. The main part of the jet is inside this dense ridge.
The density of the main part of the jet at $z=4\,r_0$ is
$\rho_{jet} \sim 10^{-3} \rho_0$ but still larger than the halo density.
Figure \ref{fig2.01} ({\it right}) shows the volume rendered image 
of the density
distribution (color scale) and the three-dimensional structure of magnetic
field lines (solid curves). The magnetic field lines are highly twisted
due to the rotation of the disk.

\subsection{Acceleration Force of Jets}
Figure \ref{press} shows the distribution of gas pressure (gray scale),
magnetic field lines (white curves), and velocity vectors for 
model $I$ ($left$), model $R$ ($middle$), and model $RC$ ($right$).
In the ideal MHD model (Model $I$), gas pressure is small in the inner 
region ($0.2 < r < 0.7$ and $0 < z < 0.1$), where magnetic 
pressure supports the disk. Mass accretion proceeds along the surface channel
where the angular momentum of the infalling gas is magnetically extracted.
Since the magnetic fields are frozen to the plasma, the mass accretion
deforms the magnetic field lines. Magnetocentrifugal force accelerates
the plasma along magnetic field lines which have 
sufficiently large angle from the rotation axis.

In resistive models (model R and model RC), magnetic field lines
are not deformed so much as those in the ideal MHD model because
matter can traverse the magnetic field lines.
Since the infalling matter loses less angular momentum than
that in the ideal MHD model, centrifugally supported inner disk is formed.
Due to adiabatic compression, gas pressure increases in the equatorial
region. Gas pressure in the surface region of the disk is larger
in model $R$ than in model $RC$ because of Joule heating.

Figure \ref{fig2.2} ({\it right}) shows the dependence of 
the inclination angle
$\theta$ of the poloidal magnetic field line in model $R$ 
depicted by a white curve in the left panel.
The angle from the rotation
axis of the disk does not exceed $30^{\circ}$.
This does not indicate that magnetocentrifugal acceleration is unimportant 
because fluid elements away from the equatorial plane 
can be accelerated by 
the magnetocentrifugal force even when $\theta < 30^{\circ}$.

Figure \ref{fig3} ({\it left}) 
shows the isocontours of temperature (gray scale contour),
magnetic field lines (white curves), and velocity vectors (arrows) for
model $R$ at $t=13.1$. 
Cold disk matter are accelerated
along the magnetic field lines and form jets as we already showed in 
\citet{kuwa00}. Figure \ref{fig3} ({\it right}) 
shows the distribution of pressure
(gray scale contour), poloidal stream lines (white curves), 
and poloidal Lorentz force
vectors 
$(\mbox{\boldmath $J$}\times\mbox{\boldmath $B$})_{\rm P}/\rho=
\left[ (\mbox{\boldmath $J$}\times\mbox{\boldmath $B$})_{r},\,
(\mbox{\boldmath $J$}\times\mbox{\boldmath $B$})_{z}\right]/\rho$
(arrows), 
where $\mbox{\boldmath $J$}=c\nabla\times\mbox{\boldmath $B$}/(4 \pi )$.
The softened gravitational potential inside $r=R_{\rm in}$ has small influence
on the formation of jets emanating from the outer radius.
The streamlines shown in Figure \ref{fig3} ({\it right})
smoothly pass through $R=R_{\rm in}$ and slowly accretes to the central object. 

The poloidal Lorentz force is almost perpendicular to the poloidal
stream lines in the launching region of the jet. 
This means that Lorentz force
collimates the outflow toward the rotation axis but acceleration
along the poloidal magnetic field lines is small.
We would like to point out that inside the disk, the
poloidal Lorentz force points toward the equatorial plane because the magnetic
field mainly has $+\phi$ component and the electric current mainly has
$-r$ component. Thus the Lorentz force compresses the torus in the vertical
direction. This force suppresses the outflow from this region. However, in the
surface layer of the disk where the magnetic field lines change their
direction from radial to vertical, the poloidal Lorentz force
changes its direction and enables outflows.

A white circle in Figure \ref{fig3} ({\it right}) 
denotes the position of the test particle
initially inside the torus at ($r$, $z$)=($1.5\,r_0$, $0.73\,r_0$)
and later
accelerated along the magnetic field line.
For this test particle, poloidal Lorentz force almost has no contribution
to the acceleration along the magnetic field line until $t\sim 13.0$.
On the other hand,
the test particle is in the region where pressure gradient force 
lifts the particle in the vertical direction.

Figure \ref{fig4} ({\it left}) 
shows the trajectories of the Lagrangian particles which 
are initially on a magnetic field line at $r=1.5\,r_0$ 
in model $R$
The initial position of particles
is selected such that they form main part of the jet.
Figure \ref{fig4} ({\it right}) shows the time
variation of the forces per mass along the streamline exerted 
on the particle which is shown
as open circles in Figure \ref{fig4} ({\it left}).
The forces accelerate the particle along its stream line
when the sign is plus and decelerate the particle when the sign is minus.
In Figure \ref{fig4} ({\it right}), 
the curve $F_c$ shows the centrifugal force 
$(v^2_{\phi}/r)_{\|}$, 
the curve $F_g$ shows the gravity $(-GM/r^2)_{\|}$,
the curve $F_p$ shows the pressure gradient force $(-\nabla P)_{\|}/\rho$,
the curve $F_L$ shows the poloidal Lorentz force per mass
$(\mbox{\boldmath$J$}\times\mbox{\boldmath$B$})_{{\rm p}\|}/\rho$,
and the curve $F_t$ shows the total force per mass obtained 
by summing the four forces.
Here the subscript
"${\|}$" means the component parallel or anti-parallel to 
the direction of the velocity vector of the particle.
The two vertical broken lines show the time when $V_z$ changes from negative
to positive ({\it left line}) and when $V_r$ changes from negative to positive 
({\it right line}).
Around $t\sim 12.5$,
centrifugal force decelerates the radial inflow.
The pressure gradient force contributes to turn the direction of motion
toward the vertical direction. When the gravitational force becomes zero
($t \sim 12.8$), the pressure gradient force along the streamline becomes
maximum, while the centrifugal force still decelerates the particle.
At $t=13.16$, centrifugal barrier finally turns the direction of the motion
of the particle to $+r$ direction. 
After the radial velocity becomes positive, centrifugal force and Lorentz
force slowly accelerate the particle along the magnetic field line 
($t>13.4$).

Figure \ref{fig6} ({\it left}) shows the
trajectories of particles which are initially located on a
magnetic field line at $r=0.8\,r_0$ for model $I$.
The initial position of particles
is selected like that the particles form main part of the jet.
Figure \ref{fig6} ({\it right}) shows the forces along the 
streamline of the particle denoted by white circles 
in Figure \ref{fig6} ({\it left}).
Until $t\sim 4.6$, centrifugal force almost balances with the radial
gravity (i.e., Keplerian rotation). Around $t=4.85$ when the particle turns 
its direction to $+r$ direction, the Lorentz force has the largest 
contribution to the acceleration of the particle.
After $t=4.9$, the summation of centrifugal force and the Lorentz force
accelerates the particle. 
This result is consistent with that of 
\citet*{kudoh98} and \citet{kato02}.
Since the magnetic forces are much larger
than those in model $R$ (Fig. \ref{fig4}), acceleration
is larger.

Figure \ref{forcerz} shows the time dependence of $r$ and $z$ components of
each force per mass acting on the Lagrangian particle shown 
as a non-filled circle
in Figure \ref{fig4} ({\it left}) for model $R$.
The left panels show the $r$-component, and
the right panels show the $z$-component.
In the radial direction, centrifugal force exceeds the radial gravity
around $t\sim 12$.
Subsequently, radial component of the Lorentz force collimates the outflow
($t\sim 14$).
In the vertical direction, the fluid element is lifted almost hydrostatically
near the ejection time ($t\sim 12.5$) and subsequently accelerated 
by the Lorentz force ($t\sim 14$).
It means that centrifugal barrier turns the direction of inflows.
At the same time, vertical component of the pressure gradient force
lifts the particle toward $z$-direction and finally it is accelerated by
the Lorentz force.
Thus the radial inflow turns into the outflow along the magnetic field line.

Figure \ref{PandPsi} ({\it top}) shows the magnetic pressure 
$\log_{10}P_B$
and the gas pressure $\log_{10}P_g$ along a magnetic field line depicted
in Figure \ref{fig2.2} ({\it left}). 
The vertical dotted line in {\it top, middle}, and {\it bottom}
shows the ejection point of the jet where $V_r=0$.
At the ejection point, the gas pressure is dominant.
On the other hand, the magnetic pressure becomes dominant in $z > 0.35$.
Figure \ref{PandPsi} ({\it middle}) shows the effective potential
\begin{equation}
\Psi_{\rm eff}= -\frac{1}{(r^2+z^2)^{1/2}}-\frac{1}{2}\Omega^2r^2,
\end{equation}
along a magnetic field line shown by a white curve 
in Figure \ref{fig2.2} ({\it left}). 
The filled circle shows the slow magnetosonic point where $V_p=V_{\rm slow}$.
Here $V_p$ is the poloidal speed and $V_{\rm slow}$ is the slow magnetosonic 
speed defined as follows,
\begin{equation}
V_{\rm slow}=\frac{1}{2}\left[ V_{\rm s}^2+V_A^2
-\sqrt{\left( V_A^2+V_{\rm s}^2\right)^2-4V_{p,A}^2V_{\rm s}^2},
\right],
\end{equation}
where $V_{p,A}^2=B_p^2/(4\pi\rho)$.

The fluid elements in the region of $d\Psi_{\rm eff}/dl<0$ are accelerated 
by the magnetocentrifugal force and those in the region of 
$d\Psi_{\rm eff}/dl>0$ accrete in the case of ideal-MHD \citep{kudoh98}
where $l$ is the line element of a magnetic field line.
However, in the resistive model, the gas in the region 
$d\Psi_{\rm eff}/dl>0$ is
ejected because the gradient of gas pressure
along a magnetic field line lifts the fluid elements and
enable them to jump over the potential barrier.
Subsequently, the gas is accelerated by magnetocentrifugal force.
On the other hand, in ideal-MHD case,
the ejection point locates at almost top of the
effective potential ($d\Psi_{\rm eff}/dl\sim 0$) and 
the slow magnetosonic point almost coincides with this point
(Fig. \ref{PandPsi} {\it bottom}).

\subsection{Quasi Stationality of Outflow}
Figure \ref{flxalB} ({\it top})
shows the time evolution of fluxes of angular momentum and mass,
\begin{equation}
Fl = 2\int_0^{2.5} 2\pi r
\left( r\rho V_z V_\phi-\frac{rB_zB_\phi}{4\pi} \right)\,dr,
\end{equation}
\begin{equation}
Fm = 2\int_0^{2.5} 2\pi r \rho V_z\,dr,
\end{equation}
at $z=3\,r_0$. 
The angular momentum and mass fluxes approach constant values around $t=25$.
Figure \ref{flxalB} ({\it bottom})
shows the distribution of the following quantities 
on a magnetic field line depicted in
Figure \ref{fig2.2}.
\begin{equation}
K = 4\pi\rho \frac{|\mbox{\boldmath $V$}_p|}{|\mbox{\boldmath $B$}_p|},
\end{equation}
\begin{equation}
\Lambda = rV_\phi -\frac{rB_\phi}{K},
\end{equation}
\begin{equation}
\Omega = \frac{V_\phi}{r}-\frac{KB_\phi}{4\pi\rho r},
\end{equation}
\begin{equation}
S = \log\left(\frac{P}{\rho^\gamma}\right),
\end{equation}
\begin{eqnarray}
E = \frac{\gamma P}{(\gamma -1)\rho}+\psi-\frac{r^2\Omega^2}{2}
   +\frac{\mbox{\boldmath $V$}_p^2}{2}+\frac{r^2(V_\phi /r-\Omega)^2}{2}.
\end{eqnarray}
In ideal MHD case, these quantities should be constant along a
magnetic field line \citep{usty99}.
The vertical dotted line shows the position where $V_r$ is zero
(ejection point of the jet).
The quantities are nearly constant even in the resistive model.
These results indicate quasi-stationarity of the outflow.

\subsection{Powers of Jets}

To study the energetics of the jet formation 
and compare them with the results by
\citet{casse04}, we compute the energy liberated by accretion $P_{\rm ACC}$
and the power of jet $P_{\rm JET}$ defined as follows:
\begin{equation}
P_{\rm ACC} = P_{\rm MEC}+P_{\rm ENT}+P_{\rm MHD}, \label{eq:13}
\end{equation}
\begin{eqnarray}
P_{\rm MEC}\!\!\!\!&=&\!\!\!\!-2\int_0^{0.3} 2\pi r\, \rho V_r
             \left(\frac{V^2}{2}+\psi\right)\,dz, \label{eq:14}\\
P_{\rm ENT}\!\!\!\!&=&\!\!\!\!-2\int_0^{0.3} 2\pi r\, \rho V_r
             \left(\frac{\gamma}{\gamma-1}\frac{P}{\rho}\right)\,dz, 
             \label{eq:15}\\
P_{\rm MHD}\!\!\!\!&=&\!\!\!\!-2\int_0^{0.3} 2\pi r\,
             \frac{c}{4\pi}\left(\mbox{\boldmath $E\times B$}\right)_r\,dz,
             \label{eq:16}
\end{eqnarray}
at $r=0.4\,r_0$, and
\begin{equation}
P_{\rm JET} = P_{\rm MEC,J}+P_{\rm ENT,J}+P_{\rm MHD,J}, \label{eq:17}
\end{equation}
\begin{eqnarray}
P_{\rm MEC,J}\!\!\!\!&=&\!\!\!\!2\int_0^{2.5} 2\pi r\, \rho V_z
             \left(\frac{V^2}{2}+\psi\right)\,dr, \label{eq:18}\\
P_{\rm ENT,J}\!\!\!\!&=&\!\!\!\!2\int_0^{2.5} 2\pi r\, \rho V_z
             \left(\frac{\gamma}{\gamma-1}\frac{P}{\rho}\right)\,dr, 
             \label{eq:19}\\
P_{\rm MHD,J}\!\!\!\!&=&\!\!\!\!2\int_0^{2.5} 2\pi r\,
             \frac{c}{4\pi}\left(\mbox{\boldmath $E\times B$}\right)_z\,dr,
             \label{eq:20}
\end{eqnarray}
at $z=3\,r_0$.

Figure \ref{jpower} shows the time evolution of the energy fluxes in the jet,
a mechanical power $P_{\rm MEC,J}$, enthalpy flux $P_{\rm ENT,J}$, 
Poynting flux $P_{\rm MHD,J}$, and the ratio $R_{\rm POW,J}$ of each power to 
the total power $P_{\rm JET}$ at $z=3$. 
The dominant energy flux at this height is the Poynting flux whose ratio
to the total power is $R_{\rm MHD,J} \sim 70\,\%$. The enthalpy flux 
and the mechanical flux contribute to $R_{\rm ENG,J} \sim 20\,\%$, 
and $R_{\rm MEC,J} \sim 10\,\%$, respectively.

\citet{kudoh97} showed that the dominant energy of a jet depends on the
strength of magnetic field. 
When the poloidal component of magnetic field is $B_p\propto r^{-2}$,
the fast magnetosonic point appears
far from the Alfv\'en point and the dominant energy of the jet is 
Poynting flux. 
In our simulations, initial magnetic field is uniform.
In such models, the fast magnetosonic 
point locates far from the Alfv\'en point \citep{kuwa00}.

Figure \ref{sigma} ({\it top}) shows the magnetic flux 
$B_p\Sigma$ 
along a magnetic field line depicted in Figure \ref{fig2.2},
$\Sigma\, (\Sigma\propto r^2)$
is a cross section of the flux tube.
Since $B_p\Sigma\sim {\rm constant}$, $B_p\propto r^{-2}$.
Figure \ref{sigma} ({\it bottom}) shows the total specific energy
$E=V_p^2/2+V_{\phi}^2/2+[\gamma /(\gamma -1)]P/\rho 
+\psi-r\Omega^{'}B_{\phi}/(4\pi\lambda^{'} )$. 
Here $V_p^2/2$ is the poloidal kinetic energy, 
and $-r\Omega^{'}B_{\phi}/(4\pi\lambda^{'})$ is the Poynting flux
divided by $\rho V_p$ where $\Omega^{'}=-V_pB_{\phi}/B_p+V_{\phi}/r$,
and $\lambda^{'}=\rho V_p/B_p$.
Figure \ref{sigma} shows that Poynting flux is dominant.
It is consistent with the case of $B_p \propto r^{-2}$
in \citet{kudoh97}.

Figure \ref{power} shows the time evolution of the mechanical power
$P_{\rm MEC}$, enthalpy flux $P_{\rm ENT}$, 
and Poynting flux 
$P_{\rm MHD}$ transported inward through $r=0.4\,r_0$ in the disk,
and the ratio of the total power of the jet 
$P_{\rm JET}$ to the total power liberated by accretion 
$P_{\rm ACC}$.
Numerical results of model $R$ indicate that 
$-P_{\rm ENT}/P_{\rm MEC} \sim 0.4$,
$P_{\rm MHD}/P_{\rm MEC} \sim 0.1$, and $P_{\rm ACC}/P_{\rm MEC} \sim 0.7$. 
Meanwhile $P_{\rm JET} \sim P_{\rm ACC}$. 
Thus, about $70\%$ of the mechanical energy released by
mass accretion powers the jet.




\section{SUMMARY AND DISCUSSION}
In this paper, we studied the acceleration mechanism of 
resistive MHD jets launched
from accretion disks threaded by weak large scale poloidal magnetic fields.
We re-computed the models we reported in \citet{kuwa00} by applying the
CIP-MOCCT scheme modified for resistive MHD equations. 
We carried out simulations for time scale about 50 inner orbital time.
Figure \ref{entp} shows the ejection mechanism of resistive MHD jets.
In mildly resistive disks, the disk gas infalls without much losing its
angular momentum because magnetic field lines are less deformed than
those in ideal MHD model. Thus, the matter accreting from the initial torus
hits the centrifugal barrier and forms a high-pressure inner disk whose 
pressure gradient force enables the accreting material to jump over the 
barrier of the effective potential.
The material lifted up by pressure 
gradient force passes through the slow magnetosonic point and is accelerated
by the magnetocentrifugal force. Mass accretion/outflow takes place 
continuously and the system approaches a quasi-steady inflow-outflow state.

The main acceleration force is the magnetocentrifugal force 
for both resistive and non-resistive models.
The acceleration point by the magnetocentrifugal force depends on
the resistivity and it
is $z\sim 0.6\,r_0$ 
when $\bar{\eta}\sim0.01$ and $z\sim 0.25\,r_0$ when $\bar{\eta}=0$.
The acceleration force for resistive model is only $25\%$ of the acceleration
for non-resistive case. We confirmed that 
Joule heating is not essential for the formation of jets.

Through resistive MHD simulations treating the accretion disk as the fixed
boundary, \citet{fendt02} showed that the jet velocity increases with 
increasing diffusivity. This result is consistent with the result of
\citet{kuwa00} that the mass outflow rate increases with the resistivity
up to some critical value. The time scale approaching the quasi-steady state
is shorter in our approach which includes the disk in simulation region
because accretion of the disk material deforms the magnetic field lines.

\citet{casse04} showed by MHD simulations assuming the magnetic diffusivity
localized inside the disk that resistive MHD jets are formed from a thin disk
threaded by global poloidal magnetic fields.
Although the initial conditions and model parameters of our simulations are
quite different from those of them, the density distribution and structures
of magnetic field lines of the disk-jet system at the final quasi-steady state
is similar \citep[see Fig. \ref{fig2.1} of ][]{casse04}.

The energy flux of jets in our mildly resistive model 
is mainly transported by the Poynting flux, while the mechanical flux
dominates the outflow in \citet{casse04}.
\citet{kudoh97} showed that the dominant energy of a jet depends on the
strength of magnetic field. 
When the poloidal component of magnetic field is $B_p\propto r^{-2}$,
the fast magnetosonic point appears
far from the Alfv\'en point and the dominant energy of the jet is 
Poynting flux. 
When $B_p\propto r^{-(2+a)}$
where $a>0$,
the fast magnetosonic point locates near the Alfv\'en point and
the dominant energy is the kinetic energy.
In our model, $B_p$ is constant with radius at the initial state.
Numerical results are consistent with those in \citet{kudoh97}
when $B_p$ decreases slowly with radius. Since magnetic energy
decreases faster in \citet{casse04}, the fast magnetosonic point
locates closer to the ejection point of the jet. Thus, in
\citet{casse04}, the jet is dominated by kinetic energy.

The jet power ($P_{\rm JET}$) 
is comparable to the energy released by mass accretion ($P_{\rm ACC}$),
which is about $70\%$ of the released mechanical energy $P_{\rm MEC}$. 
As \citet{casse04} pointed out, in ADAF 
(Advection Dominated Accretion Flows) , $P_{\rm MEC} \sim -P_{\rm ENT}$ so
that $P_{\rm ACC} \sim 0$. Thus, the resistive inflow-outflow configuration is
different from ADAF. Our numerical results can be interpreted as 
magnetic analogue of advection dominated inflow-outflow solutions (ADIOS) which 
appears in hydrodynamical models with high phenomenological viscosity 
\citep{bla99}.

\acknowledgments

We thank Dr. Seiichi Kato for helping us to extract forces acting on particles. 
This work was supported by 
the Japan Society for the Promotion of Science
Japan-UK Cooperation Science Program (principal investigators: K. Shibata 
and N. O. Weiss),
the Astronomical Data Analysis Center (ADAC) of the National Astronomical 
Observatory, Japan (NAOJ),
ACT-JST of Japan Science and Technology Corporation, and
National Science Council, Taiwan, Republic of China under the grants
NSC-90-2811-M-008-020, NSC-91-2112-M-008-006.
The numerical computations were carried out on VPP5000 
at the ADAC of NAOJ under the projects rmh17b, and rhn31b.

\clearpage


\begin{figure}
\caption{{\it Left}: Lagrangian test particles initially located 
         near the surface of 
         the accretion disk along a magnetic field line.
         {\it Right}: Schematic picture showing how to extract the force along
         the streamline.
         \label{fig2}}
\end{figure}

\clearpage

\begin{figure}
\caption{Time evolution of temperature distribution (color contours), magnetic 
         field lines (white curves), and velocity vectors (white arrows).
         {\it Top panels}: model $R$ (with Joule heating).
         {\it Bottom panels}: model $RC$ (without Joule heating).
         \label{fig2.1}}
\end{figure}

\clearpage

\begin{figure}
\caption{{\it Left}: Density distribution (color contours), velocity vectors
         (white arrows), and magnetic field lines (white curves) at $t=25$.
         Yellow arrow shows the unit velocity. 
         {\it Right}: Three-dimensional distribution of density (color)
         and magnetic field lines (blue curves).
         \label{fig2.01}}
\end{figure}

\clearpage
\begin{figure}
\caption{Pressure distribution in Model $I$ ({\it left}),
         Model $R$ ({\it center}), and Model $RC$ ({\it right}).
         White curves show the magnetic field lines, arrows
         show the velocity vectors and an arrow at upper right corner
         shows the unit velocity.
         \label{press}}
\end{figure}

\clearpage
\begin{figure}
\caption{{\it Left}: Temperature distribution at $t=20.0$
         in model $R$ ($\bar{\eta}=1.25\times 10^{-2}$).
         {\it Right}: Spatial variation of the angle $\theta$ 
         between the poloidal magnetic 
         field line depicted in the left panel
         and the rotational axis as a function of $z$.
         Broken line shows the critical angle over which the 
         magneto-centrifugal acceleration drives outflows.
         \label{fig2.2}}
\end{figure}

\clearpage

\begin{figure}
\caption{Close up of the jet launching region of model $R$ 
         ($\bar{\eta}=1.25\times 10^{-2}$)
         at $t=13.1$. 
         {\it Left}: Gray scale contour shows the temperature distribution,
         the arrows show the poloidal velocity vectors
         (an arrow at upper right corner shows the reference 
          velocity vector $=V_{\rm K0}$)
         and white curves show poloidal
         magnetic field lines.
         White circle shows the position of the Lagrangian
         particle at this time. 
         {\it Right}: Gray scale contour shows the pressure distribution,
         arrows show the poloidal component of the Lorentz force
         (an arrow at upper right corner shows the reference vector
          $=3GM/r_0^2$)
         and white curves show poloidal stream lines.
         White circle shows the position of the Lagrangian
         particle at this time.
         \label{fig3}}
\end{figure}

\begin{figure}
\caption{{\it Left}: Trajectories of Lagrangian particles 
         in model $R$ ($\bar{\eta}=1.25\times 10^{-2}$).
         Dashed curves show
         the trajectories of Lagrangian test particles. 
         Solid curve shows the
         initial disk surface. Circles are the location of test particles
         at times denoted in the figure.
         {\it Right}: Forces along a streamline of the particle shown by
         open circles in the left panel. Forces accelerate the 
         particle when the sign is plus and decelerate the particle when
         the sign is minus. Radial velocity $V_r$ and vertical velocity
         $V_z$ change their sign at the time denoted by vertical broken lines.
         Curves $F_c$, $F_g$, $F_p$, $F_L$, and $F_t$ show
         centrifugal force, gravity, pressure gradient force,
         poloidal Lorentz force, and total force.
\label{fig4}}
\end{figure}

\clearpage

\begin{figure}
\caption{{\it Left}: Trajectories of Lagrangian particles 
         in model $I$ ($\bar{\eta}=0.0$).
         Dashed curves show the trajectories of test
         particles. Solid curve shows the initial disk surface.
         Circles are the location of test particles at times denoted
         in the figure.
         {\it Right}: Forces along a streamline of the particle shown by
         open circles in the left panel.
         Forces accelerate the particle when the sign is plus and
         decelerate the particle when the sign is minus. Radial velocity
         $V_r$ and vertical velocity $V_z$ change their sign at the time
         denoted by vertical broken lines. 
         Curves $F_c$, $F_g$, $F_p$, $F_L$, and $F_t$
         show centrifugal force, gravity, pressure gradient force,
         poloidal Lorentz force, and total force.
         \label{fig6}}
\end{figure}

\clearpage

\begin{figure}
\caption{Time evolution of $r$- and $z$-components of each force
         acting on Lagrangian particle (non-filled circle) 
         in Figure \ref{fig4} (model $R$).\label{forcerz}}
\end{figure}

\clearpage
\begin{figure}
\caption{
         {\it Top}: Distribution of magnetic pressure $P_B$
         and gas pressure $P_g$ at $t=20$
         on a magnetic field line depicted in Figure \ref{fig2.2}.
         {\it Middle}: Distribution of Effective potential at $t=20$
         on a magnetic field line depicted in Figure \ref{fig2.2}.
         {\it Bottom}: Distribution of effective potential at $t=5.5$
         along a magnetic 
         field line in main part of the ideal MHD jet (Model $I$).
         Vertical dotted line shows the ejection point of the jet ($V_r>0$).
         Filled circles show the slow magnetosonic point.
         \label{PandPsi}}
\end{figure}

\clearpage
\begin{figure}
\caption{{\it Top}: Time evolution of fluxes of angular momentum $Fl$ and
         mass $Fm$. 
         {\it Bottom}: Distribution of variables, 
         $K=4\pi\rho | \mbox{\boldmath $V$}_p | / |\mbox{\boldmath $B$}_p|$,
         $\Lambda=rV_\phi -rB_\phi/K$, $\Omega=V_\phi /r-KB_\phi/(4\pi\rho r)$,
         $S=\log (P/\rho^\gamma)$, and 
         $E=[\gamma /(\gamma -1)]P/\rho+\psi-\Omega^2r^2/2+V_p^2/2
           +(V_\phi/r-\Omega)^2r^2/2$ at $t=20$
         on a magnetic field line depicted in Figure \ref{fig2.2}.
         Vertical dotted line shows the ejection point of the jet ($V_r>0$).
         \label{flxalB}}
\end{figure}

\clearpage
\begin{figure}
\caption{Time evolution of powers accelerating jet,
         mechanical power $P_{\rm MEC,J}$, enthalpy $P_{\rm ENT,J}$,
         Poynting flux $P_{\rm MHD,J}$, and the ratio $R_{\rm POW,J}$ of
         each power to total power. $R_{\rm MHD,J}$, $R_{\rm ENT,J}$,
         and $R_{\rm MEC,J}$ are the ratio of Poynting flux, enthalpy, and
         mechanical power to total power of jet. 
         \label{jpower}}
\end{figure}

\clearpage
\begin{figure}
\caption{
         {\it Top}: Magnetic flux $B_p\Sigma$ at $t=20$ along a magnetic field
                    line depicted in Figure \ref{fig2.2}.
                    $\Sigma (\Sigma\propto r^2)$ is a cross section
                    of the flux tube.
         {\it Bottom}: Specific energies at $t=20$ along a magnetic field line
                       in Figure \ref{fig2.2}. Solid curve shows the
                       total energy $E=V_p^2/2+V_{\phi}^2/2
                       +[\gamma /(\gamma -1)]P/\rho+\psi
                       -r\Omega^{'}B_{\phi}/(4\pi\lambda^{'} )$, dashed
                       curve shows the poloidal kinetic energy $V_p^2/2$,
                       and the dash-dotted curve shows the Poynting flux
                       divided by $\rho V_p$,
                       $-r\Omega^{'}B_{\phi}/(4\pi\lambda^{'})$.
          \label{sigma}}
\end{figure}

\clearpage
\begin{figure}
\caption{Time evolution of powers liberated by accretion,
         mechanical power $P_{\rm MEC}$, enthalpy $P_{\rm ENT}$,
         Poynting flux $P_{\rm MHD}$ transported inward through $r=0.4\,r_0$
         in the disk, and ratio of total power of jet
         to total power of accretion $P_{\rm JET}/|P_{\rm ACC}|$,
         $P_{\rm ACC}=P_{\rm MEC}+P_{\rm ENT}+P_{\rm MHD}$.
         \label{power}}
\end{figure}

\clearpage
\begin{figure}
\caption{Schematic picture of the ejection mechanism of the outflow
         in model $R$ ($\bar{\eta}=1.25\times 10^{-2}$).
         Gray region shows the high pressure inner torus created by
         the adiabatic compression of gases.
         \label{entp}}
\end{figure}






\clearpage
\begin{deluxetable}{lcclc}
\tabletypesize{\scriptsize}
\tablecaption{Model Parameters \label{tbl:1}}
\tablewidth{0pt}
\tablehead{
\colhead{Model} &
\colhead{$\bar{\eta}$} &
\colhead{$R_{\rm m0}$} &
\colhead{Outflow Type} &
\colhead{joule heating 
         in energy eq.}
}
\startdata
$I\dotfill$ & $0.0$ & $\infty$ & Episodic & no \\
$R\dotfill$ & $1.25\times10^{-2}$ & $1.8$ & Quasi--steady & yes \\
$RC\dotfill$ & $1.25\times10^{-2}$ & $1.8$ & Quasi--steady & no \\
\enddata
\tablecomments{In all models,
$E_{\rm th}=V^2_{s0}/(\gamma V_{\rm K0}^2)=5\times 10^{-2}$, 
$E_{\rm mg}=V_{\rm A0}^2/V_{\rm K0}^2=5\times 10^{-4}$ 
and $\rho_{\rm h}/\rho_0=10^{-3}$}
\end{deluxetable}




\end{document}